\documentclass[twocolumn,superscriptaddress,reprint]{revtex4-2}

\usepackage{amsmath}
\usepackage{amssymb}
\usepackage[dvipdfmx]{graphicx}
\usepackage{color}
\usepackage{bm}
\usepackage{enumerate}
\usepackage{siunitx}

\begin{document}

\title{Modulation of superconducting properties by the charge density wave at the surface of 2$H$-NbSe$_2$}

\author{T.~Hanaguri}
\email{hanaguri@riken.jp}
\affiliation{RIKEN Center for Emergent Matter Science, Wako, Saitama 351-0198, Japan}

\date{\today}

\begin{abstract}
{
To investigate the interplay between charge density wave (CDW) and superconductivity, we performed ultralow-temperature spectroscopic-imaging scanning tunneling microscopy on the cleaved surface of the layered superconductor 2$H$-NbSe$_2$.
We found that the superconducting-gap spectrum exhibits intricate structures reflecting the anisotropic gaps opening on multiple Fermi surfaces.
Notably, none of the characteristic energy scales apparent in the spectral gap show appreciable spatial variations, suggesting that the finite-momentum pairing is negligible.
Instead, the spectral weight near the coherence peak is modulated with the same periodicity as the CDW.
The maximum position of the coherence-peak-weight modulation coincides with neither the peak nor the bottom of the CDW modulation; rather, it aligns with the center of one of the two inequivalent triangular plaquettes that comprise the CDW unit cell.
This distribution pattern of Bogoliubov quasiparticles directly results from the broken in-plane inversion symmetry at the surface of 2$H$-NbSe$_2$, which may activate Ising spin-orbit coupling.
}
\end{abstract}

\maketitle

\section{Introduction}

Superconducting properties can develop superstructures in real space for various reasons.
An Abrikosov vortex lattice in a type-II superconductor is a classic example, where the superconducting order parameter periodically diminishes at the cores of quantized vortices.
More exotic spatially oscillating superconducting states emerge if Cooper pairs with a finite momentum can form.
The Fulde-Ferrell-Larkin-Ovchinnikov~\cite{Fulde1964PR,Larkin1965JETP,Matsuda2007JPSJ} and pair-density-wave~\cite{Agterberg2020ARCMP} states have been theoretically proposed as potential examples of such states, yet they remain elusive experimentally.

In addition to these emergent states, superconducting properties can naturally exhibit supermodulations when superconductivity coexists with another order that breaks translational symmetry of the underlying crystal lattice, such as a charge density wave (CDW).
The periodic potential of the CDW with a wavevector $\mathbf{Q}$ can lead to finite-momentum Cooper pairing between states with momenta $\mathbf{k}+\mathbf{Q}$ and $-\mathbf{k}$, etc. along with the ordinary zero-momentum pairing between $\mathbf{k}$ and $-\mathbf{k}$~\cite{Machida1981PRB}.

Among various materials in which CDW and superconductivity coexist~\cite{Manzeli2017NRM,Uchida2021JPSJ,Ortiz2020PRL}, the layered transition metal dichalcogenide 2$H$-NbSe$_2$ serves as a prominent example.
This material exhibits a $3\times3$ CDW order below \qty{30}{K} and becomes superconducting below \qty{7}{K}.
While 2$H$-NbSe$_2$ has been the subject of extensive research for decades,
the spectroscopic signature of the CDW gap remains under discussion~\cite{Kiss2007NatPhys,Rahn2012PRB,Soumyanarayanan2013PNAS,Arguello2014PRB,Pasztor2021NatCommun,Kundu2024CM}.
In addition, it is only recently that the interplay between CDW and superconductivity can be directly imaged in real space using techniques based on scanning tunneling microscopy (STM).
Liu \textit{et al.} conducted Josephson STM on 2$H$-NbSe$_2$, which utilizes a superconducting niobium tip, enabling the direct mapping of the spatial distribution of the Cooper pair density~\cite{Liu2021Science}.
They revealed that the pair density spatially oscillates with the same periodicity as the CDW modulation, whereas there is a $2\pi/3$ phase difference between the pair density and CDW modulations.

Other important quantities that should be visualized in real space include the superconducting-gap amplitude and the local density of states (LDOS) that reflect the weights of Bogoliubov quasiparticles.
These quantities can be investigated using conventional spectroscopic-imaging (SI) STM that measures the differential conductance spectrum $dI(\mathbf{r},eV)/dV$ at every pixel of the topographic STM image.
Here, $\mathbf{r}$ is the location at the surface, $e$ is the elementary charge, $I$ is the tunneling current, and $V$ is the bias voltage applied to the sample.
Cao \textit{et al.} performed SI-STM on 2$H$-NbSe$_2$ and reported that the gap amplitude is apparently modulated in-phase with the CDW modulation, while there is a $2\pi/3$ phase difference between the modulations of the coherence-peak height and CDW~\cite{Cao2024NatCommun}.

In principle, the real-space superconducting properties are linked to the structures of the superconducting gap in momentum space.
Since 2$H$-NbSe$_2$ is a multi-orbital and multi-band superconductor with multiple Fermi surfaces, each hosting an anisotropic superconducting gap~\cite{Kiss2007NatPhys,Rahn2012PRB}, the superconducting gap spectrum shows intra-gap fine structures that reflect the momentum-space variation~\cite{Guillamon2008PRB,Dvir2018NatCommun,Sanna2022npjQM}.
The spatial variations of these intra-gap fine structures are indispensable to understand  the relationship between CDW and superconductivity.
However, because the superconducting gap amplitude of 2$H$-NbSe$_2$ is only about \qty{1}{\milli\eV}, the SI-STM experiment must be conducted with a high energy resolution to resolve the fine structures.
To achieve this, we performed SI-STM at an ultra-low temperature, reaching an effective energy resolution of \qty{36}{\micro\eV}.
This high resolution enabled us to capture the subtle details of the intra-gap structures and their spatial variations.

Before we delve into the details of the experiment, let us first review the general aspects of the CDW at the surface of 2$H$-NbSe$_2$.
The crystal structure of 2$H$-NbSe$_2$, shown in Fig.~1(a), belongs to the space group $P6_3/mmc$, which is inversion symmetric.
However, each NbSe$_2$ layer, including the topmost layer observed using STM, breaks in-plane inversion symmetry [Fig.~1(b)].
The $3\times3$ CDW at the surface exhibits two domains, where the CDW maximum is located either at the anion (selenium) site or at the hollow site, as illustrated in Figs.~1(c) and 1(d), respectively~\cite{Sanna2022npjQM,Gye2019PRL,Yoshizawa2024PRL}.
In each domain, the CDW unit cell is composed of two triangular plaquettes with distinct internal atomic arrangements, reflecting the broken in-plane inversion symmetry.
We label these plaquettes as PAC1 and PAC2 for the anion-centered domain and PHC1 and PHC2 for the hollow-centered domain as depicted in Figs.~1(c) and 1(d), respectively.
\begin{figure}
\centering
\includegraphics[width=0.4\textwidth]{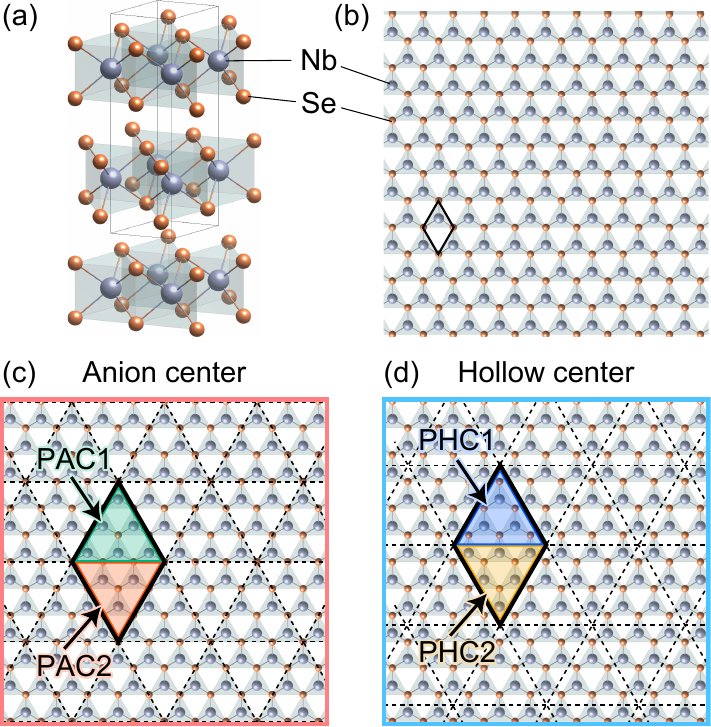}
\caption{
(a)~The crystal structure of 2$H$-NbSe$_2$ visualized using \texttt{VESTA}~\cite{Momma2011JAC}.
A rhomboid prism (thin black lines) indicates the unit cell.
(b)~Top-down view of a single NbSe$_2$ layer that breaks in-plane inversion symmetry.
A unit cell is shown by a thick black rhombus.
(c) and (d)~Top views of NbSe$_2$ monolayers exhibiting anion-centered and hollow-centered $3\times3$ CDW orders, respectively.
Black dashed lines connect the CDW maxima.
Thick black rhombuses represent CDW unit cells, which consist of two inequivalent triangular plaquettes: PAC1 and PAC2 for the anion-centered CDW, and PHC1 and PHC2 for the hollow-centered CDW.
}
\label{Fig1}
\end{figure}

It is important to note that the phase difference of $2\pi/3$ between the maxima of the Cooper pair density and CDW modulations observed by Josephson STM~\cite{Liu2021Science} means that the Cooper pairs tend to occupy one of the two inequivalent plaquettes in each domain.
Since the $2\pi/3$ phase difference has been found consistently across a large area that includes both anion-centered and hollow-centered domains~\cite{Liu2021Science}, the preferred plaquettes should be PAC1 and PHC1 or PAC2 and PHC2.
This suggests that spatial modulations of superconducting properties at the surface of bulk-inversion-symmetric 2$H$-NbSe$_2$ are influenced by the broken in-plane inversion symmetry of the monolayer.
In the following sections, we will detail the spectroscopic features within a CDW unit cell at the surface and discuss their relevance to the Cooper pair density modulation.

\section{Methods}

SI-STM experiments were conducted using a home-built dilution-refrigerator-based ultra-low temperature ultra-high vacuum (UHV) STM system similar to the setup described in Ref.~\onlinecite{Machida2018RSI}.
The lowest effective electron temperature was estimated to be $\sim\qty{120}{mK}$ by fitting the superconducting gap spectrum of aluminum to the so-called Maki function~\cite{Maki1964PTP}.
The corresponding effective energy resolution is \qty{36}{\micro eV}.
All of the data shown below were taken with this condition.

Single crystalline 2$H$-NbSe$_2$ samples were grown using the chemical vapor transport technique with iodine as the transport agent.
The sample was cleaved at liquid nitrogen temperature in a UHV environment ($\sim \qty{E-10}{Torr}$) to obtain a clean and flat (001) surface for SI-STM experiments.
Immediately after the cleaving, the sample was transferred to the STM unit, which was maintained at low temperatures below \qty{10}{K}.
An electrochemically etched tungsten wire was used as the scanning tip, which was cleaned by field evaporation using a field-ion microscope and conditioned by repeated indentations into the clean Cu(111) surface.
All these processes were performed within the same UHV chamber to avoid contamination from air exposure.

We performed all the experiments in the constant current mode with the feedback set-point tunneling current $I_\mathrm{s} = \qty{500}{pA}$ at the set-point bias voltage $V_\mathrm{s} = \qty[retain-explicit-plus]{+5}{mV}$.
Tunneling spectra were measured using a standard lock-in technique with a modulation amplitude $V_\mathrm{mod} = \qty{14}{\micro V_{\mathrm{rms}}}$ at a modulation frequency of \qty{617.3}{Hz}.
The data were collected using a commercial STM controller (Nanonis) and data analysis and visualization were performed using \texttt{SIDAM}~\cite{SIDAM}.

\section{Results}

Figure~2(a) depicts a typical STM topograph taken in a field of view that includes both anion-centered and hollow-centered CDW domains.
Zoom-in images of atomic and CDW corrugations in these domains are presented in Figs.~2(b) and 2(c), respectively.
The $dI(\mathbf{r},eV)/dV$ spectra spatially averaged over these zoom-in regions are shown in Figs.~2(d) and 2(e).
We do not find any noticeable differences between the averaged spectra taken in the anion-centered and hollow-centered domains.
The superconducting gap fully opens between $\sim\pm\qty{0.5}{mV}$, and coherence peaks develop at $\sim\pm\qty{1.2}{mV}$.
Unlike the superconducting gap spectrum of conventional isotropic superconductors, multiple fine structures are observed in 2$H$-NbSe$_2$, indicating the multi-band character and the superconducting-gap anisotropy.
Notably, there is a sharp peak at $\pm\qty{1.14}{mV}$ along with two broad humps at $\pm\qty{0.72}{mV}$ and $\pm\qty{1.30}{mV}$.
We have also inspected the second derivative of $dI/dV$, namely $d^3I/dV^3$, in which even faint features can still be distinguished.
As shown in Figs.~2(d) and 2(e), this process reveals an additional feature at $\pm\qty{1.04}{mV}$.
Fine structures can be seen above about \qty{1.4}{mV} on both positive and negative bias sides.
These structures may be extrinsic in origin, arising from the tip density of states or the bit-transition artifacts in the digital-to-analog converter for the bias voltage.
Regardless of their origin, these fine structures are sufficiently small that they do not affect the subsequent discussion.
\begin{figure}
\centering
\includegraphics[width=0.45\textwidth]{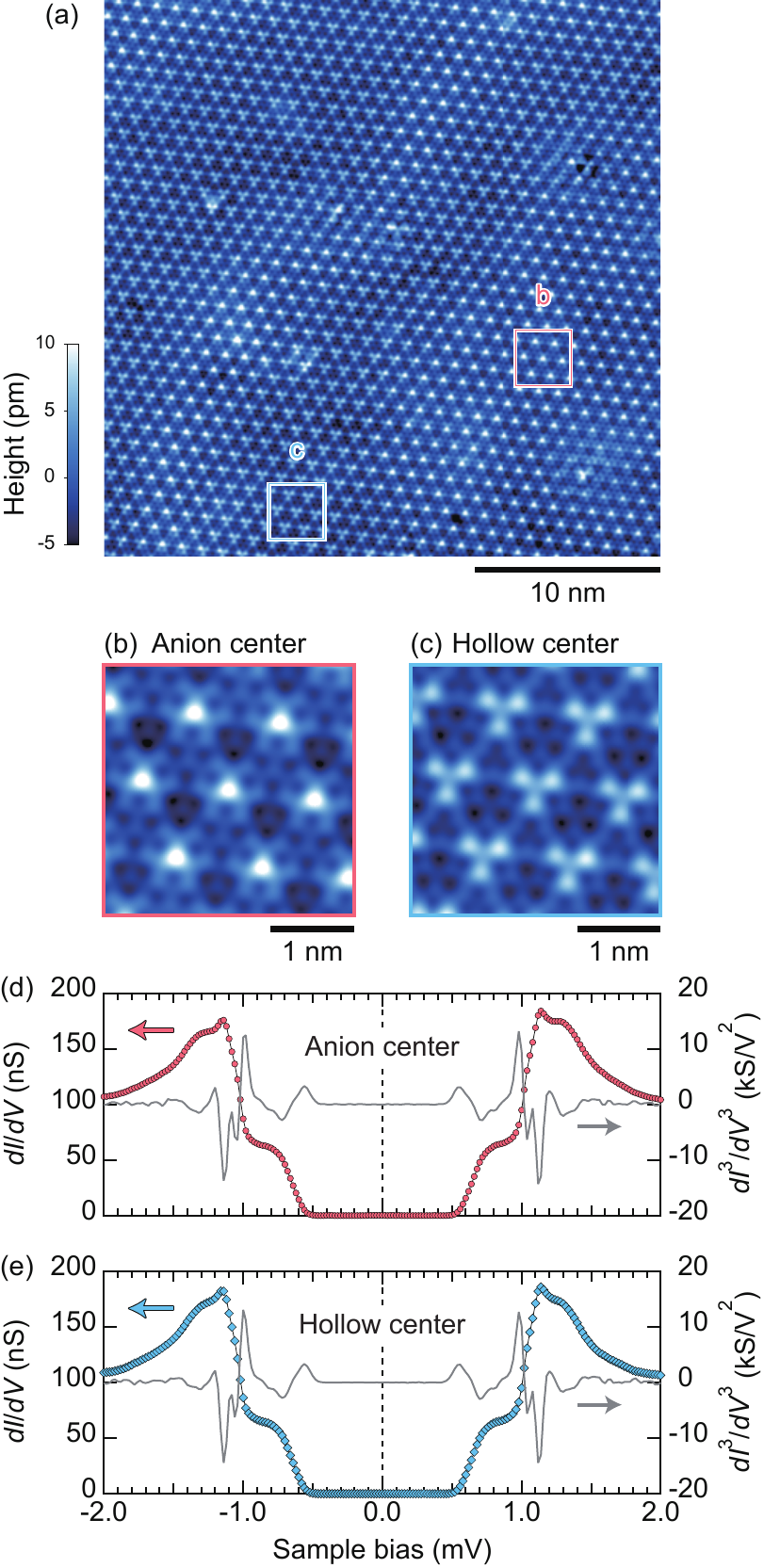}
\caption{
(a) A typical STM topograph of the cleaved surface of 2$H$-NbSe$_2$.
The feedback set-point is $I_\mathrm{s} = \qty{500}{pA}$ at $V_\mathrm{s} = \qty[retain-explicit-plus]{+5}{mV}$.
The areas highlighted in colored boxes represent zoomed-in views in (b) and (c).
(b) and (c) Zoomed-in STM topographies for the anion-centered and hollow-centered domains, respectively.
The feedback set-point condition is the same as in (a).
(d) and (e) Averaged $dI/dV$ spectra over the fields of view shown in (b) and (c), respectively.
Gray curves represent the $d^3I/dV^3$ spectra.
}
\label{Fig2}
\end{figure}

Next, we investigate the spatial variations of the energies of these characteristic fine structures in the superconducting gap.
While a faint kink at $\pm\qty{1.04}{mV}$ was difficult to identify in the individual spectrum at each pixel, we analyze the spatial variations of the other three features.
We determined the energies of these features by fitting $d^3I/dV^3$ spectra near the features with cubic polynomial functions.

Figure~3(a) presents histograms of the three characteristic energies in the anion-centered CDW area [Fig.~1(b)].
The corresponding averaged $dI/dV$ and $d^3I/dV^3$ spectra are also shown.
For clarity, we show data only from the positive bias side (empty states); however, qualitatively and quantitatively similar results were obtained for the negative bias side as well.
The full-width at half maximum of the histogram for the $\qty{1.14}{mV}$ peak is as narrow as \qty{10}{\micro eV}.
The spreads of the other two humps are also narrow, limited to several tens of \unit{\micro eV}.
The spatial distributions of the characteristic energies appear almost random, lacking any periodic patterns as depicted in Figs.~3(b)-3(d).
These results indicate that the pairing amplitude in real space is essentially uniform within our precision of a few tens of \unit{\micro eV}.
Comparably similar results are also obtained in the hollow-centered area as shown in Fig.~4.
\begin{figure*}
\centering
\includegraphics[width=0.69\textwidth]{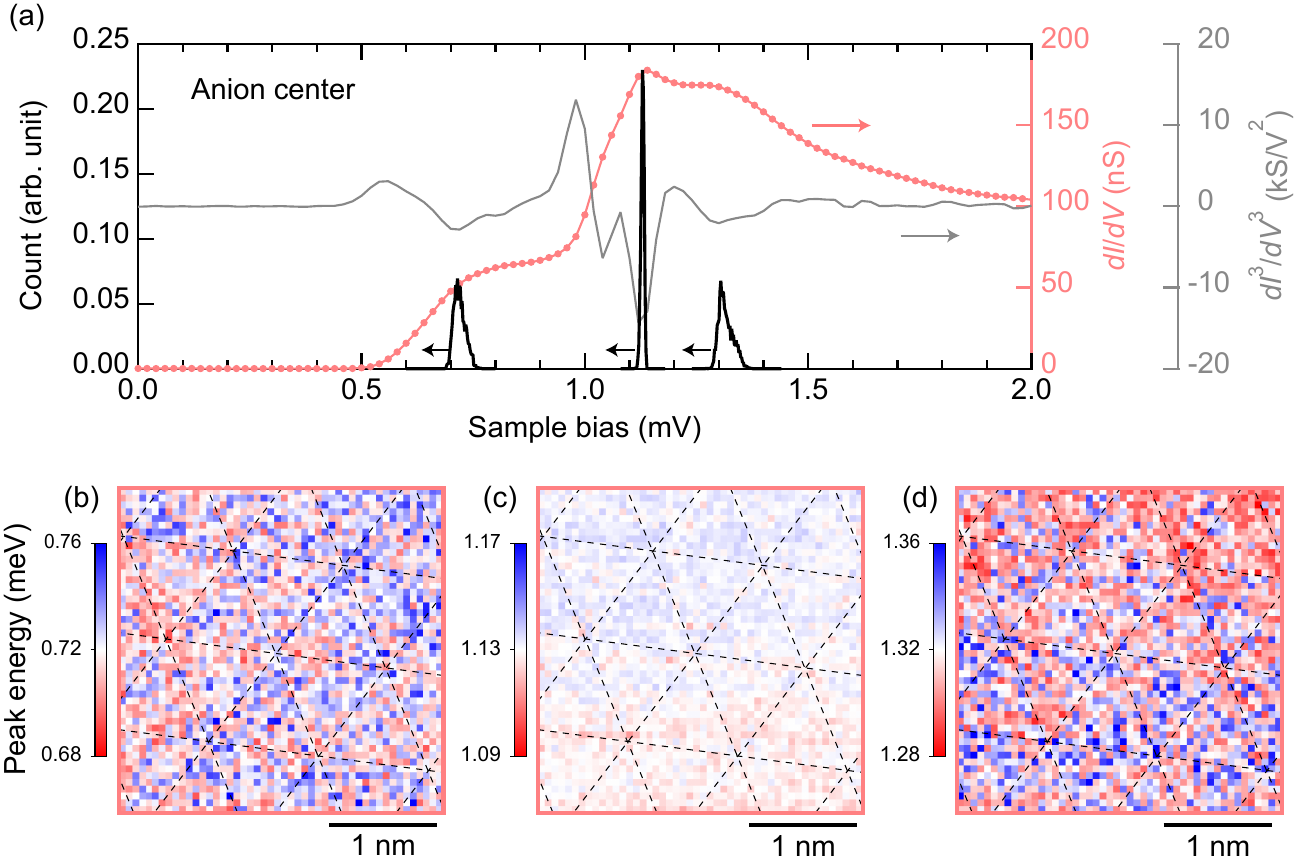}
\caption{
Spatially homogeneous superconducting energy scales in the anion-centered CDW area.
(a) Black curves represent histograms of the three characteristic energies.
The spatially averaged $dI/dV$ spectrum is depicted by a pale red curve and circles, and a gray curve shows the $d^3I/dV^3$ spectrum.
(b)-(d) Spatial maps of the three characteristic energies.
The field of view is the same as that of Fig.~1(b).
Black dashed lines connect the CDW maxima.
}
\label{Fig3}
\end{figure*}

\begin{figure*}
\centering
\includegraphics[width=0.69\textwidth]{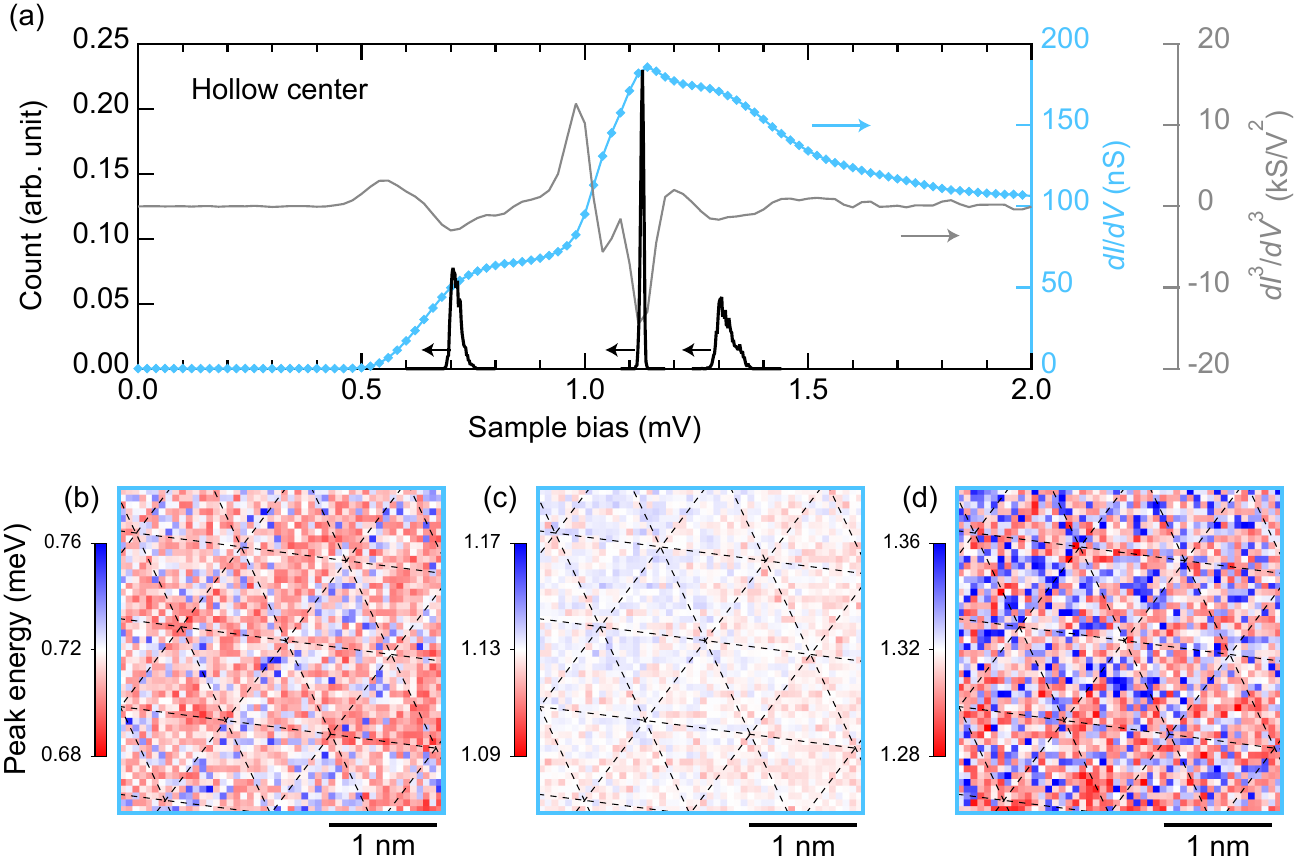}
\caption{
Spatially homogeneous superconducting energy scales in the hollow-centered CDW area.
(a) Black curves represent histograms of the three characteristic energies.
The spatially averaged $dI/dV$ spectrum is depicted by a pale blue curve and diamonds, and a gray curve shows the $d^3I/dV^3$ spectrum.
(b)-(d) Spatial maps of the three characteristic energies.
The field of view is the same as that of Fig.~1(c).
Black dashed lines connect the CDW maxima.
}
\label{Fig4}
\end{figure*}

Next, we analyze the detailed spatial and energy structures of the superconducting gap spectra.
Figure~5(a) presents tunneling spectra from the anion-centered CDW area, collected at three representative points within the CDW unit cell: the maximum positions of the CDW modulation in the topographic image and the centers of two inequivalent plaquettes labeled as PAC1 and PAC2 in Fig.~1(c).
Spectra taken from equivalent points in different CDW unit cells are shown in the same color.
The specific locations of each recorded spectrum are indicated in Fig.~5(b).
While the energies of characteristic features in the spectra do not exhibit position dependence, the relative spectral weights do vary within the CDW unit cell.
The $dI/dV$ spectra at the centers of PAC1 and PAC2 take maxima at $\pm\qty{1.14}{mV}$, while at the CDW maxima, slightly higher $dI/dV$ values tend to be observed at $\pm\qty{1.30}{mV}$.
It is important to note that if the energy resolution is insufficiently high, the multi-peak structure smears out into an apparent single peak, and the intensity variations within the fine structure could lead to an incorrect interpretation of varying gap amplitude.
\begin{figure*}
\centering
\includegraphics[width=0.8\textwidth]{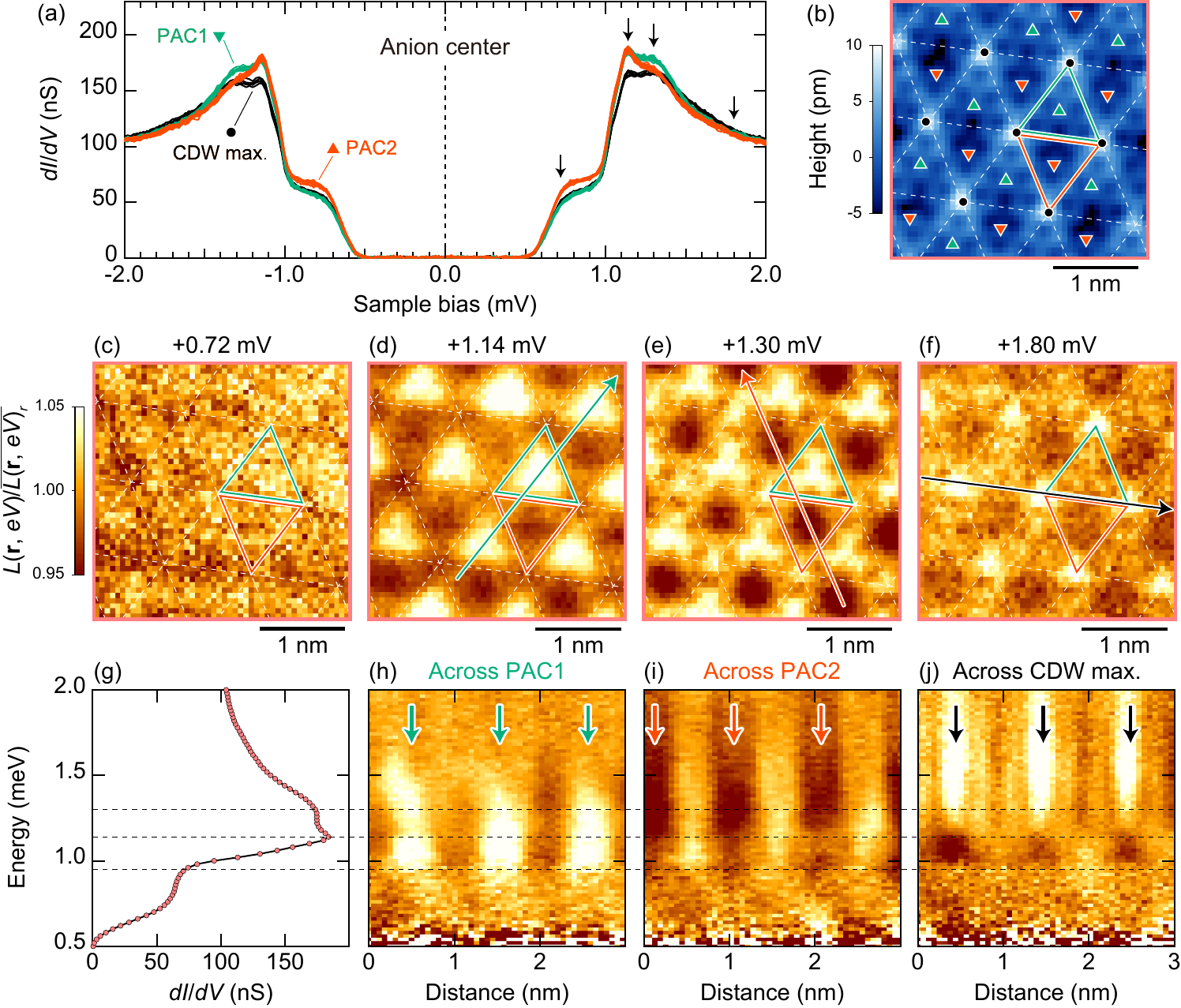}
\caption{
(a) $dI/dV$ spectra from the anion-centered area collected at three representative points within the CDW unit cell.
The arrows indicate energies of $L$ maps shown in (c)-(f).
(b) The STM topographic image showing the positions where the spectra in (a) were measured.
White dashed lines connect the CDW maxima.
The field of view matches that of Fig.~1(b).
(c)-(f) Spatial distributions of $L$ maps at characteristic energies.
Each $L$ map is normalized by its spatially averaged value to emphasize the relative contrast at a given energy.
Green and red triangles denote PAC1 and PAC2, respectively.
Arrows in (d)-(f) indicate the lines along which the line profiles shown in (h)-(j) are taken, respectively.
(g) Part of the spatially-averaged $dI/dV$ spectrum.
(h)-(j) Energy-dependent line profiles of $L$ maps along the arrows indicated in (d)-(f), respectively.
We use the same color scale as in (c)-(f).
The arrows denote the centers of PAC1 and PAC2, and the CDW maxima in the topographic image.
Horizontal dashed lines across (g)-(j) denote the characteristic energies of \qty{0.95}{meV}, \qty{1.14}{meV}, and \qty{1.30}{meV}.
}
\label{Fig5}
\end{figure*}

\begin{figure*}
\centering
\includegraphics[width=0.8\textwidth]{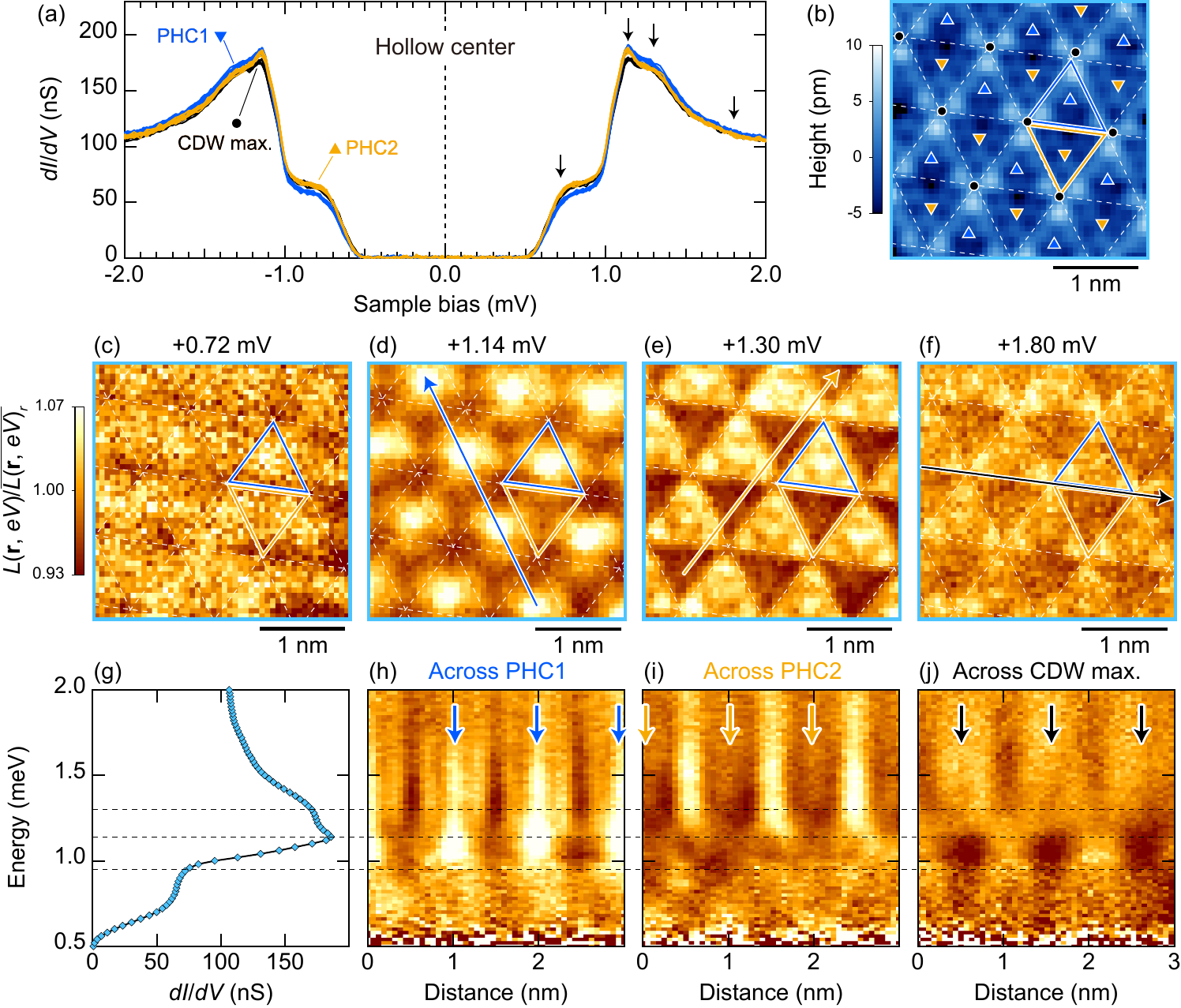}
\caption{
(a) $dI/dV$ spectra from the hollow-centered area collected at three representative points within the CDW unit cell.
The arrows indicate energies of $L$ maps shown in (c)-(f).
(b) The STM topographic image showing the positions where the spectra in (a) were measured.
White dashed lines connect the CDW maxima.
The field of view matches that of Fig.~1(c).
(c)-(f) Spatial distributions of $L$ maps at characteristic energies.
Each $L$ map is normalized by its spatially averaged value to argue the relative contrast at a given energy.
Blue and orange triangles denote PHC1 and PHC2, respectively.
Arrows in (d)-(f) indicate the lines along which the line profiles shown in (h)-(j) are taken, respectively.
(g) Part of the spatially-averaged $dI/dV$ spectrum.
(h)-(j) Energy-dependent line profiles of $L$ maps along the arrows indicated in (d)-(f), respectively.
We use the same color scale as in (c)-(f).
The downward arrows denote the centers of PHC1 and PHC2, and the CDW maxima in the topographic image.
Horizontal dashed lines across (g)-(j) denote the characteristic energies of \qty{0.95}{meV}, \qty{1.14}{meV}, and \qty{1.30}{meV}.
}
\label{Fig6}
\end{figure*}

When examining the spatial variations of LDOS from $dI/dV$ maps, care must be taken because the tip-surface distance, or effective tunneling barrier height, is changing from position to position during constant-current scanning~\cite{Kohsaka2007Science}.
This so-called set-point effect can lead to the discrepancies between $dI/dV$ and LDOS (namely Bogoliubov-quasiparticle weight) maps and can be particularly significant in the CDW system, where the STM topography may be dominated by spatial variations of the integrated LDOS up to the set-point bias voltage rather than the actual topographic corrugations.
To mitigate the set-point effect, we map out normalized conductance $L \equiv dI/dV/(I/V) = d\log{I}/d\log{V}$ instead of raw $dI/dV$.

Figures.~5(c)-5(f) display normalized $L$ maps at the three characteristic energies presented in Figs.~3, along with the one outside the coherence peak (\qty[retain-explicit-plus]{+1.80}{mV}).
Only the images on the positive bias side are shown, but the $L$ maps on the negative bias side are found to be symmetric with respect to the Fermi energy.
In the low energy region characterized by the hump at $\pm\qty{0.72}{mV}$, no clear spatial patterns are evident [Fig.~5(c)].
At higher energies, $L$ maps exhibit modulations with the same periodicity as the CDW.
Near the sharp peak in the $dI/dV$ spectra at $\pm\qty{1.14}{mV}$, $L$ exhibits a modulation that reaches its maximum at the center of PAC1 [Fig.~5(d)].
Atomic-scale modulations become visible near the $\pm\qty{1.30}{mV}$ hump [Fig.~5(e)].
At $\pm\qty{1.80}{mV}$, outside of the coherence peak, $L$ is modulated in-phase with the CDW in the topographic image [Figs.~5(b) and 5(f)].

The energy structures of the $L$ modulations can be more clearly seen in the line profiles taken along the lines across the centers of PAC1 and PAC2, and CDW maxima, as shown in Figs.~5(h), 5(i), and 5(j), respectively.
For the reference of the energy scale, the spatially averaged $dI/dV$ spectrum is also depicted in Fig.~5(g).
The CDW-related modulations emerge above \qty{0.95}{meV}, at which point $dI/dV$ abruptly increases to form a coherence peak.
The hollow-centered area demonstrates similar behaviors as depicted in Fig.~6, but a noticeable difference is observed between the line profiles across the centers of PAC2 [Fig.~5(i)] and PHC2 [Fig.~6(i)] near the onset of the CDW modulations.
The former profile resembles that across the centers of PAC1 [Fig.~5(h)], whereas the latter profile is more similar to that across the CDW maxima [Fig.~6(j)].

\section{Discussion}

Our ultra-low temperature SI-STM experiments have revealed the detailed energy and spatial structures of Bogoliubov quasiparticles in 2$H$-NbSe$_2$, which provide complementary information to the Cooper-pair-density distributions observed using Josephson STM~\cite{Liu2021Science}.
These high-resolution data serve as a benchmark for future experiments and theoretical analyses of 2$H$-NbSe$_2$.
We found that none of the energy scales characterizing the superconducting gap exhibit detectable spatial variations (Figs.~3 and 4).
The uniform energy scales of superconductivity suggest that the conventional zero-momentum pairing is dominant in 2$H$-NbSe$_2$ with a minor role of the finite-momentum component associated with the CDW~\cite{Machida1981PRB}.

In contrast to the spatially uniform superconducting gap, the spectral weights of the Bogoliubov quasiparticles exhibit pronounced energy-dependent oscillations in real space with the same periodicity as the CDW (Figs.~5 and 6).
Given the spatially uniform superconducting gap, it is reasonable to suggest that the Cooper-pair-density modulation~\cite{Liu2021Science} is directly linked to the quasiparticle-weight modulations.
This idea is corroborated by the fact that the phase difference between the quasiparticle-weight modulations near the coherence-peak energy and the CDW modulation is $2\pi/3$, which matches the phase difference observed between the Cooper-pair-density and CDW modulations~\cite{Liu2021Science}.

A more comprehensive description of the energy structures of the CDW modulations can be provided by assuming two components, A and B, in the quasiparticle weight modulations, localized at different positions and exhibiting distinct energy-dependent intensity variations: component A is maximized at PAC1 and PHC1, and component B at the CDW maxima.

The energy dependence of the weight of component A can be estimated from the line profiles across PAC1 and PHC1 shown in Fig.~5(h) and Fig.~6(h), respectively.
The weight sets in at \qty{0.95}{meV} and reaches its maximum between \qty{0.95}{meV} and \qty{1.14}{meV}.
Above \qty{1.14}{meV}, the internal structure of component A becomes evident: the weights localize at three hollow sites in PAC1 [Fig.~1(c)] and one hollow site at the center of PHC1 [Fig.~1(d)], as shown in Fig.~5(e) and Fig.~6(e), respectively.
The overall intensity of component A decreases with energy above \qty{1.14}{meV}.

Component B is concentrated near a single Se atom at the CDW maximum in the anion-centered domain [Figs.~5(e) and 5(f)], whereas in the hollow-centered domain it is distributed over three Se atoms at the CDW maximum [Figs.~6(e) and 6(f)].
The energy dependence of the weight of component B is reflected in the line profiles across the CDW maxima shown in Fig.~5(j) and Fig.~6(j). It exhibits an out-of-phase relationship with the CDW modulation in the topographic image between \qty{0.95}{meV} and \qty{1.14}{meV}, becomes in-phase and prominent above \qty{1.14}{meV}, and persists at higher energies.

This two-component model explains the different behaviors between the line profiles across the centers of PAC2 and PHC2 shown in Fig.~5(i) and Fig.~6(i), respectively.
Between the centers of PAC2, the line profile passes near one of the three hollow sites in PAC1, whereas between the centers of PHC2, near one of the three Se atoms at the CDW maximum.
Therefore, the intensities are governed by components A and B, respectively, resulting in similarities in the line profiles between PAC1 and PAC2 [Figs.~5(h) and 5(i)] and between PHC2 and the CDW maxima [Figs.~6(i) and 6(j)].

An interesting observation is that CDW-related modulations are weak below \qty{0.95}{meV}.
Angle-resolved photoemission spectroscopy experiments have revealed that the effects of CDW are significant in the Fermi surface around the K point but are hardly seen in the Fermi surface around the $\Gamma$ point~\cite{Kiss2007NatPhys,Rahn2012PRB,Kundu2024CM}.
Therefore, we speculate that the Bogoliubov quasiparticles near the coherence peak energy are dominated by the former and those below \qty{0.95}{meV} come from the latter.
Overall, however, the energy structures of the quasiparticle-weight modulations shown in Figs.~5 and 6 are complicated and require further investigations from a microscopic perspective.
We anticipate that the above phenomenological description will serve as a basis for constructing and testing future microscopic models.

We would like to point out that the observed quasiparticle-weight modulations are an intra-unit-cell phenomenon occurring within the $3 \times 3$ CDW supercell.
In other words, if we consider the supercell as the actual unit cell, there is no additional symmetry that is explicitly broken in the superconducting state.
However, intra-unit-cell modulations of superconducting properties have rarely been investigated, offering an interesting opportunity for research.
When the spatial resolution is sufficiently high, it is possible to detect such intra-unit-cell modulations in superconducting properties even in the absence of a CDW~\cite{Guillamon2008PRB,Kong2025Nature}.
For instance, Kong \textit{et al.} found an intra-unit-cell modulation in the superconducting gap amplitude in exfoliated thin flakes of iron-based superconductor FeSe$_{0.55}$Te$_{0.45}$~\cite{Kong2025Nature} and interpreted it with a model that assumes the two iron sublattices, each associated with one of the two iron atoms in the unit cell, become inequivalent.
In the case of 2$H$-NbSe$_2$, it could be possible for the two inequivalent plaquettes forming the CDW unit cell (PAC1 and PAC2, or PHC1 and PHC2) to exhibit different superconducting gap amplitudes, in principle.
However, what is actually observed is a difference in the weights of Bogoliubov quasiparticles, rather than the modulation of the superconducting gap.
It is an interesting future issue to clarify in what case the superconducting gap amplitude exhibits spatial modulations.

We emphasize that the different weights of Bogoliubov quasiparticles between PAC1 and PAC2, or PHC1 and PHC2, provide direct evidence that the broken in-plane inversion symmetry of the top-most monolayer influences the superconductivity at the surface of bulk inversion symmetric 2$H$-NbSe$_2$.
This suggests that novel phenomena, such as Ising superconductivity~\cite{Xi2016NatPhys}, may emerge at the cleaved surface due to this broken symmetry.
Additionally, when interlayer coupling is considered, the CDW stacking patterns across the surface may play a role, offering a way to modify superconductivity at the surface.
We anticipate that surface-sensitive probes, like SI-STM, will play a crucial role in exploring these novel phenomena at the 2$H$-NbSe$_2$ surface.

\begin{acknowledgments}
The author is grateful to A. Koizumi and K. Takaki for their assistance with crystal growth.
He also thanks C. J. Butler, T. Machida and M. Naritsuka for technical support and helpful discussions and comments.
This work was supported by KAKENHI Grants No. 19H05824, No. 24H00198, and No. 25H01249 from JSPS and MEXT of Japan.
This work was also supported by the RIKEN TRIP initiative (Many-body Electron Systems).
\end{acknowledgments}

\end{document}